\documentclass[prb, showpacs, twocolumn, floatfix]{revtex4}
\usepackage{bm, graphicx, amsmath, subfigure}

\newcommand{\figurewidth}{0.45\textwidth}

\begin{document}

\title{Inhomogeneous States in a Small Magnetic Disk with Single-Ion Surface Anisotropy}

\author{V.~E.~Kireev}
\email{kireev@imag.kiev.ua}

\author{B.~A.~Ivanov}
\email{bivanov@i.com.ua}

\affiliation{Institute of Magnetism NAS of Ukraine, 36-B Vernadskii avenue, 03142 Kiev, Ukraine}

\date{\today}

\begin{abstract}
We investigate analytically and numerically the ground and metastable states for easy-plane Heisenberg magnets
with single-ion surface anisotropy and disk geometry.  The configurations with two half-vortices at the
opposite points of the border are shown to be preferable for strong anisotropy.  We propose a simple
analytical description of the spin configurations for all values of a surface anisotropy. The effects of
lattice pinning leads to appearance of a set of metastable configurations.
\end{abstract}

\pacs{75.70.Rf, 75.25.+z}

\maketitle

The progress of nanotechnology permits creation of ensembles of fine magnetic particles (magnetic dots) of
nanometer scale, see for review.\cite{DemHillSlav01} Magnetic dots in the form of cylinders or prisms have
been made of soft magnetic materials like Co and permalloy\cite{Hillebrands+97, Miramond+97, Wassermann+98,
Runge+96, Cowburn+99} or highly anisotropic materials like Dy and FePt, see.\cite{Ye+95, Sun+00} Magnetic dots
and their arrays are of interest both in the basic and applied magnetism with potential applications including
high-density magnetic storage media.\cite{Smyth+91}

Usually a small magnetic particle is considered as being in the monodomain state with a homogeneous saturated
magnetization (or N\'eel vector for antiferromagnets).  During the last few years it had been established that
the distribution of magnetization within the dots made of soft magnetic materials can be quite nontrivial;
namely, various inhomogeneous states resulting from the magnetic dipole-dipole interaction appear.  In recent
years interest in such states for submicron particles has risen significantly.  A small enough non-ellipsoidal
dot exhibits a single-domain nearly uniform magnetization state, either so called \emph{flower} and
\emph{leaf} states.\cite{UsovPesch94-2, CowbWell98, Runge+96, CowbAdWell98} When increasing the size of the
dot above a critical value, vortex state occurs.\cite{UsovPesch93, FernCerj00, ShiTehrSchein00, Wassermann+98,
Pokhil+00, Shinjo+00} The main property of such states is the non-saturating value of the total dot
magnetization, nearly zero for vortex states and nearly saturated, but smaller that saturated, for leaf and
flower states.

O'Shea and coworkers\cite{OsheaPer94, JohnsPerOshea96, OsheaPer96} have observed non-satura\-ted states for
the rare-earth ferromagnetic granules with high anisotropy and the size about of 5\,nm.  A possible
explanation of this fact is that these particles are in non-uniform states.\footnote{An alternative
explanation based on the hypothesis of the presence of a random single-ion bulk anisotropy have been done by
the authors of\cite{OsheaPer94, JohnsPerOshea96, OsheaPer96}.}  On the other hand, it is clear that the
concepts of non-uniform states referred to above and caused by a magnetic dipole interaction cannot be applied
directly to such small particles made with highly-anisotropic material.  In this concern, some other sources
of non-uniformity need to be found.

The appearance of non-uniform states for small \emph{bcc} atomic clusters with taking into account the
\emph{single-ion} surface anisotropy have been shown numerically by Dimitrov and Wysin.\cite{DimWysin94-1,
DimWysin94-2} Garanin and Kachkachi in the resent work\cite{GarKachk03} investigated the effective anisotropy
caused by such a non-uniform spin distribution for small magnetic particles. The difference of the properties
of the spins on the surface and in bulk could be considered as a defect destroying the homogeneity of a
sample.  It is clear that due to the surface a homogeneous ordering is distorted or even broken.

In real magnets the surface could produce the surface anisotropy for two reasons.  First, the main origin of
magnetic anisotropy can be caused by the anisotropy of spin-spin interactions (the case of \emph{exchange}
anisotropy).  For this case even on an ideal atomically smooth surface the spins have different coordination
numbers than in bulk, and consequently the intensity of the exchange interaction changes.  For the surface
exchange anisotropy the direction of the chosen axis is the same as in bulk and has no connection to the
surface.  This effect could lead to the non-uniform states in some special cases only, mostly in the presence
of an external magnetic field, for example the surface spin-flop transition,\cite{Mills68, Wang+94} and the
states caused by the magnetic field for easy-axial ferromagnets.\cite{IvVolkMerk02} Second, in real magnets
surface atoms have a different environmental symmetry.  Thus, the surface distorts a crystalline field that
acts on a magnetic ion, and the anisotropy is changed drastically.  It leads to a specific \emph{single-ion}
surface anisotropy for the spins with a preferred axis coinciding with the normal to the surface.  This model
is considered by Dimitrov and Wysin for \emph{fcc} iron clusters;\cite{DimWysin94-1, DimWysin94-2} we would
like to investigate this case both analytically and numerically.  Note that the surface effects, in
particular, the surface anisotropy, have been considered by many authors,\cite{Aharoni.all, Shilov+99prb} but
in most of these works the ground states has been assumed to be homogeneous, and the surface terms are only
accounted in dynamics.  On the other hand, it is obvious that for fine magnetic particles the role of the
surface becomes much more important than for bulk materials.  The effects cased by the surface considered as a
defect are proportional to $N^{-1/3}$, and their role increases when the size of the particle tends to the
nanometer scales.

Note that similar problems arise in the other domains of condensed matter physics, where a role of surface is
important.  These are textures in liquid crystals\cite{deGennes74} and in a superfluid ${}^3$He,
see.\cite{He3} For the A-phase of ${}^3$He (${}^3$He-A) the unit vector order parameter $\bm{l}$, $\bm{l}^2 =
1$ is perpendicular to the surface of a vessel.  ${}^3$He can not be in equilibrium with its own vapor; it
fills the vessel completely at temperatures when it is superfluid ($T \lesssim 2$\,mK).  Thus, the vector
$\bm{l}$ should be perpendicular to the surface of the ${}^3$He-A sample.  The analysis shows that the order
parameter becomes non-uniform, and, moreover, it is singular for any simply-connected vessel.\cite{He3}

It is clear that such effects may be observed in all finite samples of ordered media with vector order
parameter and a strong surface anisotropy of the form $B(\bm{m} \cdot \bm{n})^2$, where $\bm{n}$ is the normal
to the surface, $\bm{m}$ is the order parameter, and $B$ is the constant of single-ion surface anisotropy,
which orients $\bm{m}$ with respect to the surface.  For the ${}^3$He-A, the boundary condition could be
described as a limit of an \emph{infinitely strong} surface anisotropy $B < 0$, $|B| \rightarrow \infty$, with
easy axis perpendicular to the surface.  The concept developed for ${}^3$He-A could be a good guide for a
theory of fine magnetic particles with surface anisotropy.  On the other hand, the situation for magnets is
more general: the magnitude of the surface anisotropy for magnets is finite, and the magnetic moment could be
inclined with respect to the axis of surface anisotropy.  As we will show below finiteness of anisotropy could
lead to the states with non-uniform spin distributions but without singularities.

The outline of the paper is as follows.  In Sec.~\ref{s:model} we discuss classical models for a small
magnetic particle supporting simplest non-uniform spin distribution, caused by surface anisotropy, which is
planar and two-dimensional ($2D$) model.  This means that the spins parallel to one plane and the spin
distribution depends effectively on only two space coordinates $(x, y)$.  Sec.~\ref{s:infinite} is devoted to
the planar continuum $2D$ model in the limit case of the infinite surface anisotropy, where exact solutions
are found and analyzed.  In Sec.~\ref{s:finite} the same model will be considered for the case of finite
anisotropy.  Sec.~\ref{s:numerics} contains results of direct numerical simulations for the $2D$ lattice
models and the consideration of pinning effects that can be estimated from the continuum model.  The analysis
of thermal and topological stability is also done in this section. The last Sec.~\ref{s:conclusion} contains
the resume of obtained results and a short discussion them in concern with other similar systems.

\section{Model} \label{s:model}

There are two approaches to the analysis of the static and dynamic properties of magnetic materials: discrete
microscopic and macroscopic.  The microscopic approach is based on a discrete spin Hamiltonian in which the
spins $\bm{S}_i$ (quantum or treated quasi-classically, as will be done below) are specified at the lattice
sites $i$.  In discrete models the magnetic anisotropy can be introduced in two different ways: as single-ion
anisotropy, and as anisotropy of the exchange interaction.  To describe them, the spin Hamiltonian is chosen
in the form
\begin{equation}\label{ham0}
\mathcal{H} =
\sum_{<i j>} J_{\alpha}S_i^\alpha S_j^\alpha +
\sum_i K_\alpha \bigl(S_i^\alpha \bigr)^2 +
\sum_{i} B_{\alpha\beta}(i) S_i^\alpha S_i^\beta \;.
\end{equation}
Here $S_i^\alpha$ is the projection of a classical spin on the symmetry axis $\alpha$ of the bulk crystal.
The summation in the first term is over all the nearest neighbors in the lattice, $J_{\alpha}$ is an
anisotropic exchange tensor.  The constant $K_\alpha$ and function $B_{\alpha\beta} (i)$ describe the volume
and the surface single-ion anisotropy energies, respectively. For the crystals with rhombic or higher
symmetry, all tensors describing volume characteristics can be diagonalized simultaneously.  The tensor
function $B_{\alpha\beta} (i)$ is nonzero only near the surface and abruptly decreases in the depth of the
sample.  The surface creates another chosen direction, a normal to it, and enters a local system of
coordinates, in which the tensor $B_{\alpha\beta} (i)$ is diagonal.  We neglect in the Hamiltonian
\eqref{ham0} a dipole-dipole coupling and a Zeeman interaction with an external magnetic field.

We shall use a simple version of \eqref{ham0} with an uniaxial symmetry for the bulk properties ($z$ as a
chosen axis, for definiteness) and with nearest neighbors interaction only:
\begin{multline}\label{ham}
\mathcal{H} =
- J\sum_{i, \delta} \bigl(S^x_{i} S^x_{i + \delta} + S^y_{i} S^y_{i + \delta} + 
\lambda S^z_{i} S^z_{i + \delta}\bigr)  \\
+ K_z \sum_i \bigl(S_i^z \bigr)^2 + B \sum_{i'} \bigl(\bm{S}_{i'}\bm{n}\bigr)^2 \;.
\end{multline}
Here $J$ is the exchange integral, $\lambda$ is the anisotropy parameter of the exchange interaction, and
$\bm{\delta}$ are the vectors of the nearest neighbors, the summation over $i'$ in the last term includes only
the surface sites, where the number of the nearest neighbors differs from the volume one.  To more adequately
compare the lattice and continuum models, we assume that the vector $\bm{n}$ is a normal to the surface, but
not a direction given by the Miller indices.

The sign of the exchange integral plays no role for the statics of non-frustrated magnets with a bipartite
lattice.  Moreover, a model without dipole-dipole coupling is more adequate for antiferromagnets than
ferromagnets. For simplicity we use below the ferromagnetic representation of spin distributions, i.e. $J >
0$.  The transition to the antiferromagnetic case for a bipartite lattice is trivial: we introduce sublattices
and change the directions of the spins in one of them.

The continuum approximation of \eqref{ham0} is based on a free energy functional $\mathcal{W}[\bm{m}]$ that
depends on the local normalized magnetization $\bm{m}(\bm{r})$, $\bm{m}^2 = 1$.  Using the standard smoothing
procedure of a lattice model, we write down the functional $\mathcal{W}[\bm{m}]$ as
\begin{multline}\label{w}
\mathcal{W}[\bm{m}] =
\frac{1}{2} \int_\Omega \frac{S d\bm{r}}{a^3} \Bigl\{J a^2\bigl[(\bm{\nabla}m_x)^2 +
(\bm{\nabla}m_y)^2 \\ 
+\lambda (\bm{\nabla}m_z)^2 \bigr]
+ K m_z ^2 + a B(\bm{m}\bm{n})^2 \delta(\bm{r}-\bm{r}_s) \Bigr \} \;.
\end{multline}
Here $\Omega$ is the volume of the particle, the vector $\bm{r}_s$ parameterizes the surface, $\delta(\bm{r})$
is the Dirac delta-function, $a$ is the lattice spacing, and $S$ is the cross-section area.  The solution of
the Euler-Lagrange equation for \eqref{w} gives spin configurations with a preferential direction close to the
surface.  It is clear that the measure of inhomogeneity depends on the problem parameters and the sample
shape.  A simple consideration shows that for the fixed shape there are only two relevant parameters.  The
first one is the characteristic radius $R / r_0$, where $r_0 = a \sqrt{J / K}$ at $\lambda = 1$ or $r_0 = 2a
\sqrt{\lambda / (1 - \lambda)}$ at $K = 0$ is the magnetic length, defined in the same way as for bulk
materials.\cite{Iv+98} The second parameter is the ratio of the exchange integral to the surface anisotropy $B
/ J$.  To simplify analysis we consider a model with a purely planar spin distribution.  Such distributions
appear for magnetic vortices at strong enough easy-axis anisotropy, $\lambda < \lambda_{c}$, where
$\lambda_{c} \sim 0.7 $ when $r_0 \simeq a$.\cite{Wysin94} In this case we obtain the one-parameter model
characterized by the ratio $B / J$.  As we will see below such a model demonstrates a wide set of
inhomogeneous states and allows complete analytical and numerical investigations.  We restrict ourselves the
case of one more simplification, namely, a model which allows $2D$ spatial spin distributions, i.e. such
distributions which depend only on two spatial variables, say $x$ and $y$.  Apparently such a simplification
is applicable to an island of a magnetic monoatomic layer shaped as a disk.  For numerical simulation we will
choose a fragment of the two-dimensional square lattice in the form of a disk.  However, applicability of
obtained results is not limited by this concrete case.  It is easy to imagine situations when the same
spatial-two-dimensional distribution is realized.  As an example one can regard a ferromagnetic particle with
the volume easy-plane anisotropy, having a form of a cylinder with the base parallel to the easy-plane (the
$xy$-plane) and with the axis along the $z$-axis.  If one considers that the surface anisotropy constant $B$
in \eqref{w} is positive then the normal to the surface is the hard axis of the surface anisotropy.  It is
clear, that any planar spin distribution with $S_z(i) = 0$ ensures both the minimum of the volume and the
surface anisotropy on the upper and bottom cylinder surfaces.  In this case non-uniformity is caused only by
the lateral cylinder surface, and one can expect that the distribution will be a spatial-two-dimensional one,
with the same character as for the purely two-dimensional problem.

\section{A strong border anisotropy in a continuum approach} \label{s:infinite}

We shall start from the simplest model to describe effects of surface anisotropy.  Consider a disk-shaped (or
cylinder-shaped, see above) magnet, with $xy$-plane as an easy-plane, and assume a $2D$ spin distribution.  We
assume that the magnetization is a two-dimensional unit vector, in a polar mapping: $m_z = 0$ and
$\bm{m}_\perp = \hat{\bm{e}}_x \cos\phi + \hat{\bm{e}}_y \sin\phi$, where $(\hat{\bm{e}}_x, \hat{\bm{e}}_y,
\hat{\bm{e}}_z)$ is the basis in the spin space and $\phi = \phi(x, y)$ is the angle between $\bm{m}$ and
$\hat{\bm{e}}_x$.  The magnetic energy of the disk takes the form:
\begin{equation}\label{w2}
\mathcal{W}[\phi] =
J S^2 \biggl[\frac{1}{2} \int_\Omega dS\,
(\bm{\nabla}\phi)^2 + b \int_\Gamma d\chi\, \cos^2(\phi-\chi)\biggr] \;.
\end{equation}
Here $\Omega$ is the area of our disk-shaped magnet with the radius $R$, the contour $\Gamma$ is the border
circle, and $(\rho, \chi)$ are the polar coordinates in the plane of magnet.  The parameter $b$ is
proportional to the constant of a border anisotropy, $b = (B/J)(R/a)$.  We choose $b \geq 0$, and the
preferential surface directions are tangent.  This choice is motivated above; one more reason is that such an
effective term can be used to model the magnetic dipole interaction.\cite{IvZasp02} The function $\phi(\rho,
\chi)$ may have singularities inside the disk $\Omega$.  Minimal configurations for the energy \eqref{w2} are
constructed from solution of the respective Euler-Lagrange equations, which is the scalar Laplace equation
\begin{equation}\label{Laplace}
\bm{\nabla}^2\phi = 0 \;,
\end{equation}
with the boundary condition at $\rho = R$
\begin{equation}\label{border}
R \frac{\partial\phi}{\partial\rho} \biggm|_{\rho=R} - b
\sin2[\phi(\rho, \chi)-\chi] = 0 \;.
\end{equation}
Thus, this is a problem with a nonlinear boundary condition.

In the absence of the boundary anisotropy, $b = 0$, homogeneous solutions $\phi = \text{const}$ satisfy
simultaneously \eqref{Laplace} and \eqref{border}, and this trivial case is not considered.  First of all, we
analyze possible solutions in the limit of strong border anisotropy, $b = \infty$, when the problem becomes
linear and can be solved exactly.  The boundary condition leads to the two possible solutions $\phi(R, \chi) =
\chi \pm \pi/2$.  Such ambiguity of the boundary conditions here differs from the classic internal Neumann
problem of mathematical physics and the relevant physics will be discussed below.  The solutions in both cases
can be constructed via harmonic functions, as well it can be done in two-dimensional electrostatics.\cite{LL7}
The general solution of the Laplace equation $\phi$ can be written via a complex potential $u(z)$ of integer
charges $q_k$ placed at the points $z_k$:
\begin{equation}\label{distr}
\phi=\text{Im} [u(z)],
\qquad u(z) = \sum_k q_k \ln(z - z_k) + \text{const} \;.
\end{equation}
These charges have a simple physical meaning, they describe well-known in-plane vortices, which have been
repeatedly discussed in regard to $2D$ magnetism.  We introduce a complex representation for the coordinate
plane $xy$, $z = x + iy$.  The functional $\mathcal{W}[u]$ is rewritten as
\begin{multline}\label{w-complex}
\mathcal{W}[u] =
J S^2 \biggl[\frac{1}{2} \int_0^R \rho d\rho \int_{|z|=\rho} \frac{d z}{i z} \biggm| \frac{d u}{d z} \biggm|^2  \\
+ \frac{b}{8} \int_{|z|=R} \frac{d z}{i z} \frac{(\xi^2-1)^2}{\xi^2} \biggr] \;,
\end{multline}
where
\begin{equation}\label{xi}
\xi^2= \frac{z^*}{z} \exp(u-u^*) \;.
\end{equation}

In the continuum approximation the energy $\mathcal{W}[u]$ is logarithmically divergent close to points where
the in-plane vortices (charges) $q_k$ are placed.  To describe these singular solutions in the continuum model
we have to introduce a cut-off parameter of the order of the lattice spacing.  Singularities cost much energy,
and one could expect that configurations with a global minimum of $\mathcal{W}[u]$ should be sought among the
nonsingular functions $u(z)$ in the area $\Omega$ or functions with a small number of singularities.

\subsection{Vortex-like configurations}

\begin{figure}
\subfigure[Centered vortex preferable for a strong border anisotropy. \label{f:vortex-c}] 
{\includegraphics[bb = 100 500 320 720]{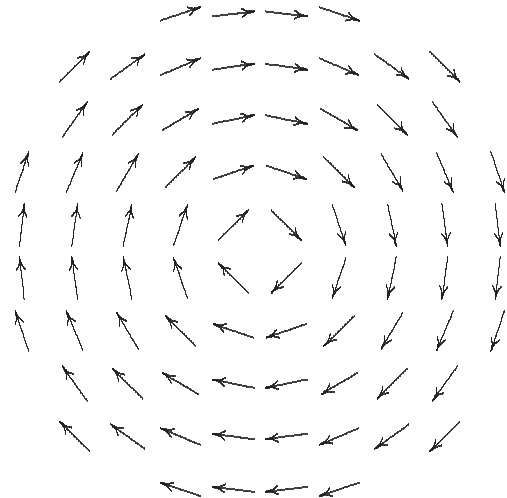}}

\subfigure[Non-centered vortex preferable for a weak border anisotropy. \label{f:vortex-d}]
{\includegraphics[bb = 100 500 320 720]{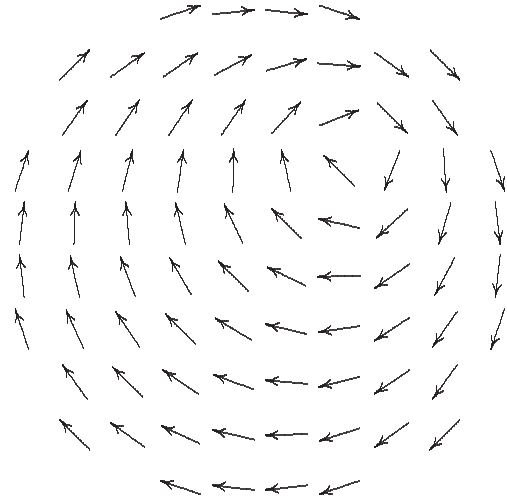}}
\caption{Numerically calculated vortex-like states for the discrete model \eqref{ham} with $\lambda = 0$, $K_z
= 0$ and $R = 5a$. \label{f:vortices}}
\end{figure}

The simplest solution with one singularity is a centered vortex, see Fig.~\ref{f:vortex-c}, generated by the
functions $u = \ln z \pm i\pi/2$ with the energy
\begin{equation}\label{e-vortex}
E_v =
J S^2 \pi\ln \biggl(\frac {R} {r_\epsilon} \biggr) \;,
\end{equation}
where $r_\epsilon$ is a cut-off parameter for vortex states of the order of the lattice spacing $a$.  Besides
these solutions the others are non-centered vortices for infinite $b$, see Fig.~\ref{f:vortex-d}, generated by
\begin{equation}\label{vortex-d}
u(z) =
\ln(z - z_0)+\ln \biggl(z - \frac {R^2} {z_0^\ast} \biggr) \pm i\frac{\pi}{2} \;,
\quad \text{where} \quad |z_0| \le R \;,
\end{equation}
with the vortex placed at the point $z_0$.  They also satisfy the conditions $\phi = \chi \pm \pi/2$ on the
border $\Gamma$.  As seen from \eqref{vortex-d}, the interaction between the vortex and the border, which may
be considered as a consequence of the boundary condition \eqref{border}, is equivalent to the coupling between
the vortex and the image vortex placed outside the disk at the inverse symmetric point respective to the
border circle.  The calculation of the energy covers only the area $\Omega$ and the singularity of the
reflected charge gives no effect.  The energy of non-centered vortex for infinite $b$ (fixed boundary
conditions $\phi = \chi \pm \pi / 2$ on the border) is given by
\begin{equation}\label{e-vortex-d}
E_v^{\text{(d)}} =
J S^2 \pi \biggl[\ln \biggl(\frac{R}{r_\epsilon} \biggr) - 
\ln \biggl(1 - \frac{|z_0|^2}{R^2} \biggr) \biggr] \;.
\end{equation}

The first term coincides to a proper energy of the vortex given by \eqref{e-vortex}, and the second term is
the energy of the interaction between the vortex and the border; it is a repulsive one. Besides it another
force acts on the vortex: the vortex has a tend to escape from a finite area in order to decrease $|\bm{\nabla
m}|$, and, thereby being attracted to the border.  In the case of $b = \infty$ the repulsive force prevails,
the vortex is stabilized at the furthest point from the border, and the second term in \eqref{e-vortex-d} is
absent.

\subsection{Configurations with two half-vortices on the border}

The above considered vortex-like distributions are some of the simplest spin distributions, minimizing surface
anisotropy, not only for the circle shape but for a border in the form of any simple contour.  Indeed, going
around a simple closed contour, the normal $\bm{n}$ to it turns to $360^{\circ}$.  This means that the
topological characteristic of the planar unit vector, so called \emph{vorticity},\cite{Iv+98} $q$ equals to
$\pm 1$ for the vector $\bm{n}$. Obviously, those magnetic vortices having the vorticity $q = 1$ are quite
probable candidates to realize the energy minimum.  Nevertheless, vortices with any $q \neq 0$ admittedly
possess singularities inside the sample.  The analysis of such distributions where magnetization has no
singularities in the bulk is of interest. A simple analysis demonstrates that in this case, as well as for
$^3$He-A singularities should appear on the border.

To explain this, consider the behavior of the vector field $\bm{m} = \hat{\bm{e}}_x\cos\phi +
\hat{\bm{e}}_y\sin\phi$ on the border circle $\Gamma$.  The boundary condition requires that the vector
$\bm{m}$ be parallel to the border. It can be presented by two ways: $\bm{m}$ may be parallel or antiparallel
to the tangent vector $\hat{\bm{\tau}} = \bm{n} \times \hat{\bm{e}}_z$.  Assuming that $\bm{m}$ is nonsingular
inside $\Omega$, the circle $\Gamma$ can be divided into an even number of alternating regions: in half of
them $\bm{m}$ has to rotate clockwise and in the others --- counterclockwise.  Thus, besides the above
considered vortex-like solutions, there exist configurations regular inside the circle $\Omega$ and with
singularities on the border, see Fig.~\ref{f:hv-2}.  (Such singularities in the three-dimensional case are
referred to \cite{He3} as vortex lines.)  The simplest two-singularity solutions can be written as
\begin{equation}\label{hv}
u(z) =
\ln(z - R e^{i\phi_1}) + \ln(z - R e^{i\phi_2}) + i\pi \pm i\frac{\pi}{2} \;.
\end{equation}

This is a field created by two charges placed at the border points $Re^{i\phi_1}$ and $Re^{i\phi_2}$.  It is
easy to check that the conditions $\phi = \chi \pm \pi/2$ are satisfied on the border wherever where $\phi(R,
\chi)$ is defined.  To calculate the energy thoroughly we have to introduce the cut-off parameter
$r_\epsilon'$ and integrate over the disk $\Omega$ except two half-circles of radius $r_\epsilon'$ centered at
the charges.  Under the condition that the cut-off regions do not overlap, $R(\phi_1 - \phi_2) \gg a$, the
energy of the configurations are
\begin{equation}\label{e-hv-d}
E_{\text{hv}} =
J S^2 \pi \biggl[\ln \biggl(\frac {R}
{r_\epsilon'} \biggr) - \ln2 - \ln\biggm| \sin\frac {\phi_1 - \phi_2} {2} \biggm| \biggr] \;.
\end{equation}
Here $r_\epsilon'$ is the corresponding cut-off parameter.  The continuum approximation does not provide a
relation between $r_\epsilon$ and $r_\epsilon'$ and we used numerical calculations for the lattice model to
find it out.  These calculations show with a good accuracy that $r_\epsilon = r_\epsilon'$, and we will assume
that in the following.  The minimum of \eqref{e-hv-d} is achieved for charges placed at the opposite points of
the border, it is given by
\begin{equation}\label{e-hv}
E_{\text{hv}}^{\text{min}} =
J S^2 \pi \biggl[\ln \biggl(\frac{R}{r_\epsilon} \biggr) - \ln 2 \biggr] \;,
\end{equation}
Thus, the interaction of surface charges with each other is also repulsive. Comparing the expressions
\eqref{e-vortex} and \eqref{e-hv}, we see that the energies for both configurations are logarithmically
diverged and differing by the constant.  Thus, the configuration with two half-vortices at the opposite points
of the border is preferable to the single vortex for $XY$-model.

\section{Finite values of a surface anisotropy} \label{s:finite}

In this section we consider the case of a finite surface anisotropy.  At $b < \infty$ the boundary condition
\eqref{border} is nonlinear.  It is easy to see that the only centered vortex from all configurations with the
vortex inside the sample is an exact solution for any finite values $b$.  A non-centered vortex is not a
solution of our problem at finite $b < \infty$.  Such states are absent in the continuum model, but they
became metastable in the discrete model because of lattice pinning.  The numerical calculations shows that
their energies depend weakly on the surface anisotropy constant.  This class will be considered in
Sec.~\ref{s:numerics}.

\begin{figure}
\subfigure[Almost homogeneous  configuration, $B / J = 0.2$. \label{f:hv-02}]
{\includegraphics[scale = 0.9, bb = 80 520 340 700]{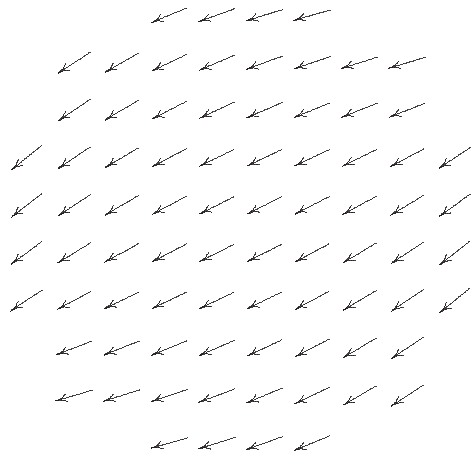}}

\subfigure[The charges are far from the border, $B / J = 0.5$. \label{f:hv-05}]
{\includegraphics[scale = 0.9, bb = 80 520 340 700]{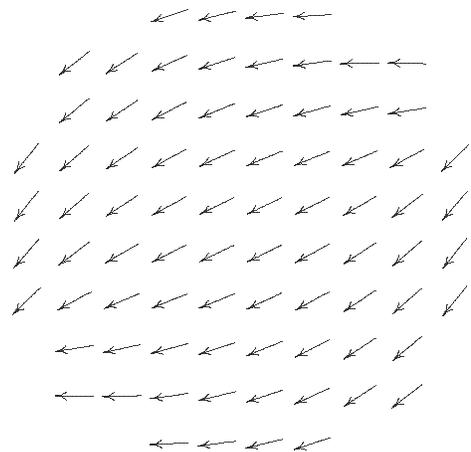}}

\subfigure[The charges are close to the border, $B / J = 2.0$. \label{f:hv-2}]
{\includegraphics[scale = 0.9, bb = 80 520 340 700]{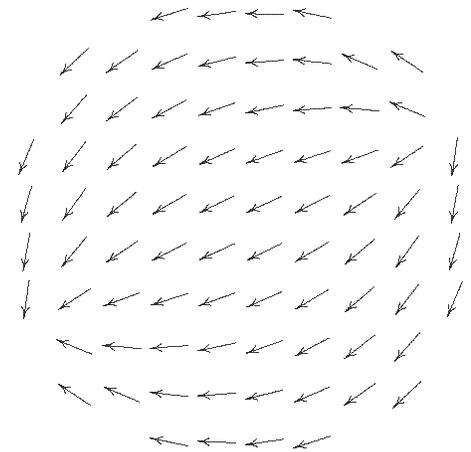}}

\caption{Minimal non-topological configurations for the discrete model \eqref{ham} with $\lambda = 0$, $K_z =
0$ and $R = 5a$ for different values of border anisotropy. \label{f:hv}}
\end{figure}

The solutions with two half-vortices on the border \eqref{hv} for finite anisotropy $b < \infty$ transforms to
non-singular solutions with two vortices placed \emph{outside} the disk at the opposite points $z_0$ and
$-z_0$, where $|z_0| > R$.  This distribution is generated by the function
\begin{equation}\label{nt}
u(z) = 
\ln(z - z_0) + \ln(z + z_0) + i\pi \pm i\pi / 2 \;.  
\end{equation}
The particular exact solutions of the problem \eqref{Laplace}, \eqref{border} with an arbitrary $b$ have been
found by Burylov and Raikher \cite{BurRaikh94} for a distribution of the vector director near the surface of a
cylindrical solid particle embedded in a monodomain nematic liquid crystal.  Using of the boundary condition
\eqref{border} for the function \eqref{nt} gives the value of $z_0$ in the form
\begin{equation}\label{alpha}
|z_0|^2 =
R^2(1 + \sqrt{1 + b^2}) / b \;.
\end{equation}
For such values $z_0$ the boundary condition \eqref{border} satisfy exactly.  The energy of the configuration
is equal to
\begin{multline}\label{e-hv-fin}
E_{\text{hv}} (b) =
J S^2 \pi \biggl[\ln\frac{1 + \sqrt{1 + b^2}}{2} + b - \frac{b^2}{1 + \sqrt{1 + b^2}} \biggr]  \\
- 2 J S^2 \int_0^{\phi_0} x\tan x\,dx \;.
\end{multline}
The latter term arises due to the cut-off close to the half-vortices which are introduced for $|z_0| - R \le
r_\epsilon$, and $\phi_0 = \arccos[(|z_0| - R) / r_\epsilon]$.  Its contribution is important for a high
enough surface anisotropy, $B \gtrsim J$ only, see Fig.~\ref{f:hv-2}.

\begin{figure}
\includegraphics[width = \figurewidth, bb = 120 500 500 720]{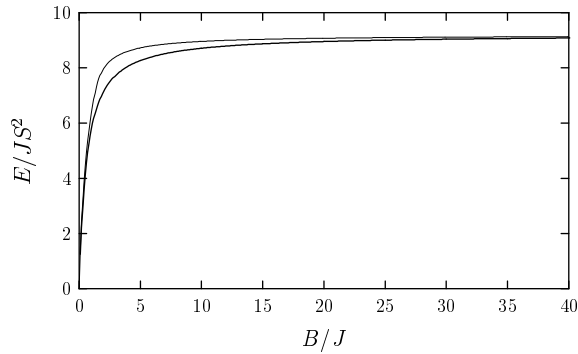}
\caption{The dependence of the energy of the minimal configuration versus the surface anisotropy for the disk
with the radius $R = 10a$.  The thin line is a two-charge approximation; the thick line is a numerically
calculated result for the lattice model.} \label{f:eb}
\end{figure}

It is easy to see that for any finite $b$ the energy of the two-charge configuration is lower than its limit
value \eqref{e-hv}, and it decreases monotonically with decreasing $b$.  Another limit case of small surface
anisotropy $b \to 0$ leads to the almost homogeneous distribution $\bm{m}$, see Fig.~\ref{f:hv-02}, with the
nearly zero energy $E = J S^2 \pi b$.  When $b$ increases, the vector field $\bm{m}$ is curved to the
diametrically pair of points, and the energy increases, see Fig.~\ref{f:hv-2}.  These features are in good
agreement with that obtained numerically for discrete finite system.  The dependency of $E_{\text{hv}}$ versus
the surface anisotropy $B$ from the Hamiltonian \eqref{ham} is plotted in Fig.~\ref{f:eb} together with that
for the continuum model \eqref{w2}.  The discrepancy of the curves is connected with the discreteness effects,
which are important for small samples, for larger system radius (the value of $R$ till $R = 30a$ has been
used).  In the case of $b \to \infty$ the vortex energy is higher than the two-charge configuration energy for
$XY$-model, and the vortex states are also metastable for any finite $b$.

\section{Numeric simulation and lattice effects} \label{s:numerics}

For our model with rather strong volume and surface anisotropy, the characteristic size is $|\bm{\nabla m}|
\sim a^{-1}$ and it is not obvious that effects of discreteness can be neglected.  An exact analysis of the
discrete model requires numerical calculations, but some qualitative results can be obtained using the lattice
potential method.  For a direct numerical simulation we basically used the $XY$-model with $\lambda = 0$,
i.e. with an extremely strong easy-plane anisotropy (some results concerning the finite $\lambda$ will be
discussed in conclusion).

\subsection{Numerical simulation}

For numerical calculation of the equilibrium states we started from the discrete Hamiltonian for the magnetic
energy \eqref{ham}.  Calculations have been performed starting from a random initial configuration or from a
configuration given by \eqref{distr} with constants $z_k$, $q$ appropriate for a considered problem.  The
energy minimization has been performed through a Seidel-like algorithm with the successive exact solution of
the local equilibrium equation for a fixed site that can be obtained from the following one-site energy
\begin{equation}
E_L =
-\bm{SH} + \frac{B}{2}(\bm{Sn})^2 - \frac{\mu}{2}\bm{S}^2 \;,
\end{equation}
where $\mu$ is a Lagrange multiplier for the condition $|\bm{S}| = 1$ and $\bm{H} = J \sum_{<>}(S^x
\hat{\bm{e}}_x + S^y \hat{\bm{e}}_y + \lambda S^z \hat{\bm{e}}_z)$ is the effective field created by the
nearest neighbors of the fixed site.  The term with $B$ is present for the border spins only.  In a simple
case $B = 0$ we obtain $\bm{S} = \bm{H} / |\bm{H}|$.  When $B \neq 0$ a more complicated analysis of the roots
of the equilibrium condition $d E_L / d\bm{S} = 0$ is needed.  Among these roots $\bm{S}_{\text{min}}$ we
choose the value that gives the deepest minimum of $E_L$.  For all minimizations we observe that this
procedure converges to one of the stable configurations.  The configuration appearing during minimization
energy process was mainly dictated by the choice of the initial configuration.

\begin{figure}
\includegraphics[width = \figurewidth, bb = 150 500 500 720]{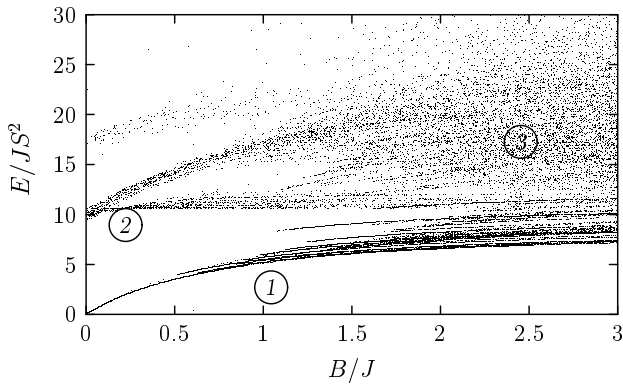}
\caption{Metastable configurations for the $R = 8a$ disk.  Shown approximately $3 \cdot 10^4$ dots.  Regions
``1'' and ``2'' are shown in Fig.~\ref{f:eb_hv} and \ref{f:eb_v}.} \label{f:eb_all}
\end{figure}

\begin{figure}
\includegraphics[width = \figurewidth, bb = 150 500 500 720]{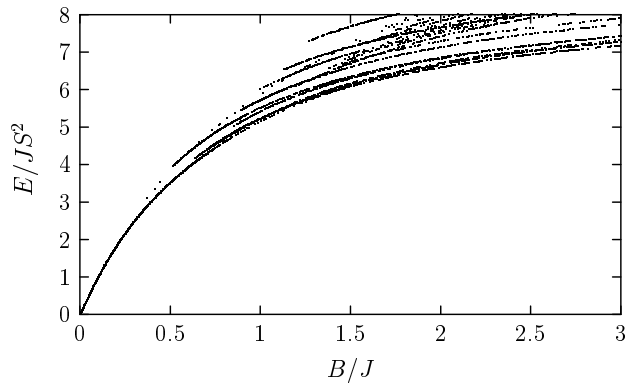}
\caption{Minima with two half-vortices for the $R = 8a$ disk.} \label{f:eb_hv}
\end{figure}

\begin{figure}
\includegraphics[width = \figurewidth, bb = 150 500 500 720]{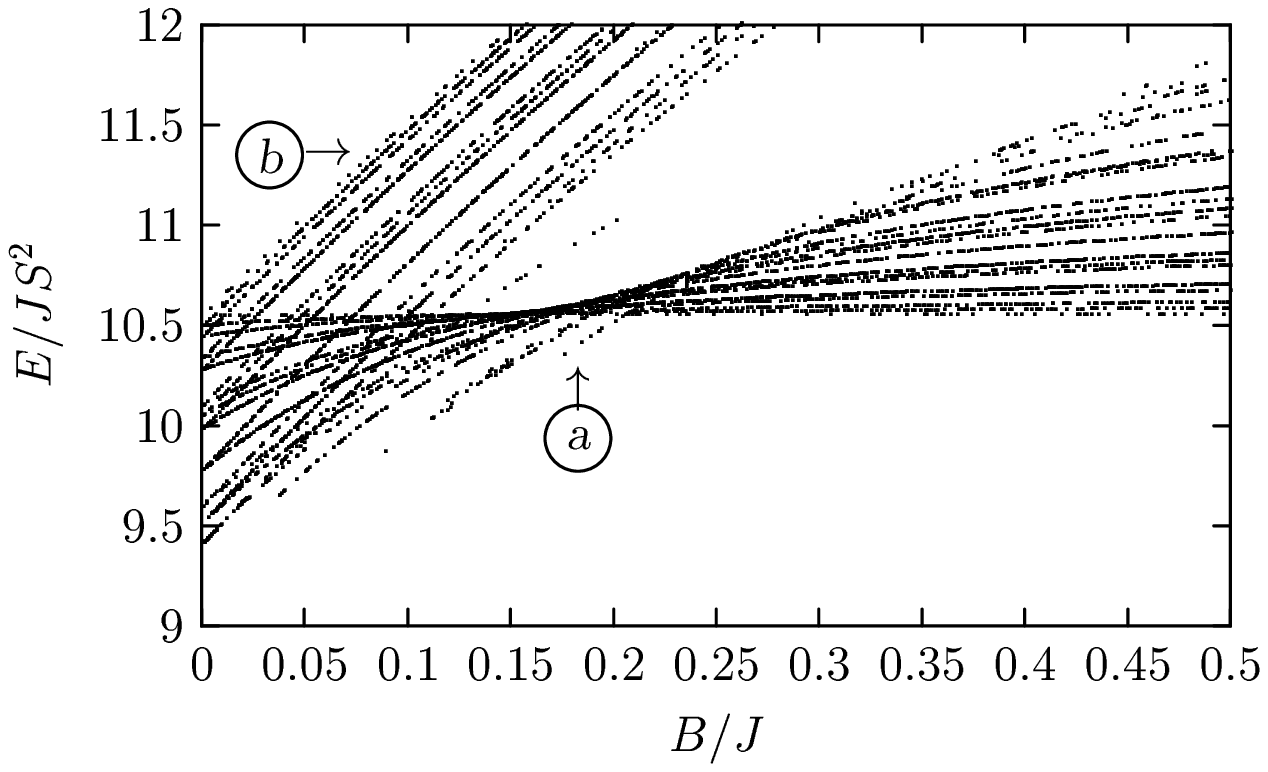}
\caption{Vortex minima close to $B = 0$ for the $R = 8a$ disk.} \label{f:eb_v}
\end{figure}

In order to explore all metastable minimal configurations in the lattice model \eqref{ham} with $\lambda = 0$
we performed more than $10^7$ minimization procedures according with the described scheme.  Initial
configurations and the surface anisotropy $B$ are chosen randomly.  The obtained energy values are presented
by dots on the plane $(B/J, E/J S^2)$, see Fig.~\ref{f:eb_all}, the system size is chosen small enough to show
a discrete nature of the possible states.  Such an analysis allows to judge both the energy absolute minimum
for a given $B/J$ and the presence of metastable states.  It is seen that in some plane regions (marked as
``1'' and ``2'') the dots are grouped in more or less well-defined lines, which obviously corresponds to the
most stable states and describes the dependence of their energies on $b$. The characteristic regions are
present in Figs. \ref{f:eb_hv}, \ref{f:eb_v}.  The region (marked as ``3''), in which the dots are distributed
practically randomly (in fact, there the dots also are fitted by lines), corresponds to high energy
states. They are not subjects of interest. To classify the spin states the positions of singularities of the
function $\phi (x, y)$ have been analyzed numerically and the positions of poles (vortices), which are placed
inside of a disk or on its border, have been obtained. Such an analysis demonstrated the presence of all
states described above, including non-centered vortices and states with non-symmetrically placed surface
singularities, but yet some less favorable states namely antivortices with the distribution like $\phi = -\chi
+ \text{const}$, where $\chi$ is the polar coordinate, instead of $\phi = \chi + \text{const}$, characteristic
for vortices. Let us discuss the obtained results.

First of all, the given analysis has confirmed that the symmetrical states with two singularities possess the
minimal energy.  In the region of the small anisotropy and energy $E \lesssim 4.0$ (here and after energy
values are presented in units of $J S^2$) only state with symmetric half-vortices are present, see details of
this region in Fig.~\ref{f:eb_hv}.  At $B/J \gtrsim 0.5$ other well-defined lines of dots appear, which also
correspond to states with two half-vortices, however with broken symmetry.  These states have higher energy
and they are unstable at small $B$, but at larger $B$ they become metastable due to surface pinning effects.
With $B$ increase first of all the states pinned in the vicinity of non-regular regions of a surface, which
result from cutting a circle specimen from the square lattice appear.  With further increase of $B/J \geq 1.5
\div 2$ the number of asymmetric states grows.

\begin{figure}
\includegraphics[width = \figurewidth, bb = 150 500 500 720]{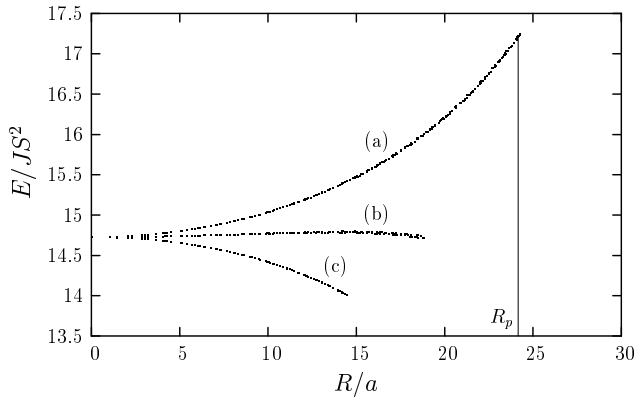}
\caption{The energy of a non-centered vortex versus its displacement for three values of surface anisotropy:
(a) $B/J = 0.4 \gg B_c$ (attraction to the center),
(b) $B/J = 0.04 \simeq B_c$ (equilibrium),
(c) $B/J = 0.004 \ll B_c$ (repulsion from the center).
Radius of the disk is $30a$. The radius of the pinning region ($R_p \sim 24$) is maximal for the largest value
of $B$ and decreases for lower values. \label{f:er}}
\end{figure}

The second interesting region of plane at the energy $E \sim 10$ corresponds to vortex states.  Its details at
small $B/J$ are depicted in Fig.~\ref{f:eb_v}.  It is worth to note that according to the analytical
consideration the centered vortex presents at any $B$ and its energy does not depend on $B$.  Besides that
state there exist non-centered vortices stabilized by the lattice pinning.  Since the state with non-centered
vortex at the finite $b$ is not an exact solution (unlike to the non-singular case), they will be analyzed
numerically in the next subsection with a simple qualitative model of pinning.  At large enough anisotropy the
energies of non-centered vortices are higher compared to the state with the centered vortex.  However, at
small anisotropy, ($B/J \leq 0.2$ for the system size $R = 8a$ used for Figs.~\ref{f:eb_all}--\ref{f:eb_v}) an
interesting effect emerges: non-centered vortices become more favorable than the centered one (the
correspondent region marked as ``a'').  This effect could be described as a change of the sign of the
effective interaction between the vortex and the border at some value $B = B_c$.  (Let us remind that the case
$B = \infty$ corresponds to fixed boundary conditions, while the case $B = 0$ corresponds to free boundary
conditions, which are associated with repulsion and attraction of the vortex to the border, respectively,
see\cite{Iv+98}.)  This effect is present also for big values of the radius, see Fig.~\ref{f:er}, in which is
plotted the vortex energy calculated in the model \eqref{ham} versus its displacement for three value of
surface anisotropy and the radius $R = 32a$.  The characteristic value of the surface anisotropy $B_c$
decreases inverse proportionally to the system size.  The values found numerically for $R = (5 \div 30) a$ can
be extrapolated by the dependence $B_c/J \sim 1.2(a/R)$.

At $B = 0$ the vortex and antivortex have the same energy and in the region of extremely small $B$
antivortices are also reliably observable, see Fig.~\ref{f:eb_v}, region ``b''.  However, when $B$ increases
the energy of antivortices grows rapidly and we do not discuss them.

\subsection{Lattice effects and vortex stability}

For non-uniform states the lattice pinning of singular points in the spin distribution both vortices and
surface singularities (half-vortices) plays an essential role.  It is interesting to discuss such points in
more details.  The continuum model neglects a discrete nature of crystals and the pinning effects.  The
simplest way to describe analytically lattice effects and, in particular, to investigate the local stability
of metastable states, is to introduce an effective periodical potential (Peierls-Nabarro potential) into the
continuum model.  Schnitzer shows,\cite{Schnitzer96} see also,\cite{Iv+98} that for in-plane vortices this
potential is independent of the values of out-of-plane anisotropy parameters (for $\lambda < 0.8$) and can be
presented in the simplest form as $U_{PN} (x, y) = \kappa J S^2\pi [\sin^2(x\pi/a) + \sin^2(y\pi/a)]$, where
the origin is chosen at the point which is equidistant from lattice sites, and the numeric parameter $\kappa
\simeq 0.200$.\cite{Schnitzer96} The potential minima are attained at all points like $\bm{r} = n\bm{e}_x +
m\bm{e}_y$, where $m$, $n$ are integers, $|\bm{e}_x| = |\bm{e}_y| = a$, and the saddle points are at $(n +
1/2)\bm{e}_x + m\bm{e}_y$ and $n\bm{e}_x + (m + 1/2)\bm{e}_y$.  A metastable state with a vortex shifted from
the center to the point $\bm{r}$ exists only when the sum $E(\bm{r}) = E_v (\bm{r}, b) + U_{PN} (\bm{r})$ has
a minimum at this point.  The loss of stability manifests itself as ruptures of lines in Fig.~\ref{f:eb_hv}
and \ref{f:eb_v}, see also Fig.~\ref{f:er}.

Then it is easy to show that the non-centered vortices are held by the pinning potential and are stable if
their coordinates are inside the circle of radius $R_p$.  The radius of the pinning region $R_p$ is determined
from the explicit expression \eqref{e-vortex-d} for the energy of the vortex placed at the point $r_0$ as
\begin{equation}\label{r-pinning}
a\frac{dE_{\text{vor}}^{\text{(d)}}(r_0)}{dr_0}\biggm|_{r_0=R_p}=
\frac{\pi\kappa}{a} \;,
\end{equation}
and the case $a \ll R$ leads to $R_p = R -  a/\kappa$.

Thus the vortices can be pinned everywhere inside the sample except the thin strip close to the border.  Their
energies relative to the zero level of the centered vortex lie in the band of the width $\sim J \ln(R/a)$.
Such states are frequently observed in numeric simulations for the discrete model when initial configurations
for the minimization are chosen randomly.

Although a detailed analysis of thermal fluctuations and decay of metastable states is beyond the scope of
this work, their role can be discussed on the basis of the previous estimates.  The above introduced $R_p$ is
the radius of the region where pinning disappears, i.e. at $r \rightarrow R_p$ the barrier height separating
states with a vortex placed in adjacent lattice sites, becomes to zero.  It is also reasonable to introduce
the function $R_p(E)$, such that at $r < R_p(E)$ the barrier height between these two states is higher than
some value $E$.  Naturally, $R_p(E) \rightarrow R_p$ at $E \rightarrow 0$, $R_p(E) \rightarrow 0$ at $E
\rightarrow E_b^{\text{max}}$, where $E_b^{\text{max}} = \kappa J S^2\pi$ is the maximal pinning energy.  For
intermediate region $ E \ll E_b^{\text{max}}$ a simple calculation yields
\begin{equation}\label{U-pinning}
R - R_p(E) =
\frac{a}{\kappa}\frac{E_b^{\text{max}}}{E_b^{\text{max}} - E} \;,
\end{equation}
and for all values of $E_b^{\text{max}} - E \sim E_b^{\text{max}}$ the value of $R_p(E)$ is again near to $R$.
Thus, the role of thermal fluctuations at $k_B T \ll E_b^{\text{max}}$ can be considered as negligently small,
and the above described metastable states may be manifest as long-lived ones even for finite temperature.  On
the other hand at $k_B T \geq E_b^{\text{max}}$ metastable states like the non-centered vortex will not be
manifest and only the centered vortex should be considered.

For two-charge configurations the lattice potential also creates others metastable configurations with higher
energies than the energy of configurations with maximally separated charges.  Their analysis is similar to the
one that has performed for the case of a non-centered vortex.  Two pinned charges on the border can be
approached only down to the angle $\phi_p = |\phi_1 - \phi_2| \simeq \pi a/\kappa R$.  Consideration of
thermal fluctuations can be done for non-centered vortices as well and it leads to the similar results,
practically all such states are metastable.

In conclusion of this section discuss the stability of vortices as a topologically nontrivial configuration
under transform to non-topological one. Inside of two topologically different classes of states --- with
vortex or with two surface singularities --- effective relaxation to the most favorable state inside of the
given class is possible.  However, the previous results show that the vortex-like configurations with the
centered vortex have higher energy than the two-charge configuration, and are metastable.  Therefore, the
state with centered vortex may relax toward the most profitable state with two surface singularities.  The
simplest scenario of the vortex decay is the following.  The vortex moves to the nearest point on the border
and its counterpart moves also to it.  The point, where they merge, is a saddle point of the path with the
energy $E_{\text{sad}} = J S^2 \pi \ln(R/\epsilon)$ over the centered vortex energy.  This state is referred
as a \textit{boojum} or \textit{fountain} in the ${}^3$He theory,\cite{He3, Mermin79} where it is a true
minimum.  Further, the merged charges decouple and move along the border: one --- in the clockwise direction
and another --- in the counterclockwise direction to the most distant positions.  Thus, the energy
$E_{\text{sad}}$ is the barrier height between the two classes of configurations, and it can be used for the
analysis of a thermal (or quantum tunneling, for low temperature) decay of vortex states.  Note, the barrier
is nothing to do with pinning potential.  Its value does not contain the parameter $\kappa$, but it is
proportional to $\ln (R/a)$ and is much higher that the exchange energy $J S^2$.  Thus, vortex states can be
stable even at high enough temperature comparing with the Curie temperature $T_c \sim J S^2$, and the
probability of the decay of the vortex state is very low even at the temperatures comparable with $T_c$.

\section{Concluding remarks} \label{s:conclusion}

A strong surface anisotropy for easy-plane Heisenberg magnets destructs the homogeneous ordering and leads to
the two types of static structures: the vortex state and the state with pair of half-vortices on the surface.
For finite anisotropy the latter state becomes non-singular.  This state is energetically favorable for all
finite values of surface anisotropy.  The energy gap between it and the vortex state is of the order of the
exchange energy, but the energy barrier is much higher that the exchange energy.  The strong bulk anisotropy
leads to well pronounced effects of lattice pinning, and large number of metastable states appears as well.

It is interesting to compare these results to those which have been obtained for fine particles made with soft
magnetic materials such as permalloy magnetic dots, where the non-uniform states are caused by the
magnetic-dipole interaction.  The common point for these cases is not only the presence of the vortex state
but also the presence of non-topological non-uniform states, leaf or flower states.\cite{UsovPesch94-2,
CowbWell98, Runge+96, CowbAdWell98} The distinction consists in the fact that for soft magnetic particles
there are non-singular vortices with the out-of-plane magnetization component while in our problem with the
strong bulk anisotropy the only in-plane vortices with a singularity are presence.  It is likely that in
virtue of this for permalloy particles there is a very much pronounced transition from the vortex state to the
non-topological one with the system size decreasing, while in our problem the vortex state is always less
favorable energetically.  It is worth to note that our preliminary numerical data indicate the appearance of
such a transition at a weak easy-plane anisotropy; an extended discussion of this problem is beyond the scope
of the present work.

It is also interesting to note that the spin distribution in the non-singular state of our $2D$ problem
resembles the distribution having axial symmetry and the plane of symmetry perpendicular to the axis obtained
by Dimitrov and Wysin \cite{DimWysin94-1, DimWysin94-2} for $3D$ particles where both the volume and surface
anisotropies are presented.  Recently, the stable three-dimensional analog of vortices, hedgehog configuration
has been discovered for a ball-shaped particle with strong normal border anisotropy by numeric
calculations.\cite{Labaye+02} On the other hand, for the superfluid $^3$He-A, which is defined in terms of our
model by use of the infinitely strong surface anisotropy and isotropic volume properties, the true minimum
constitutes less symmetric state (boojum, or fountain) with one surface singularity and without the symmetry
plane.\cite{He3, Mermin79} In our case the ''boojum-like'' distribution appears only for non-stable saddle
point, which separates the vortex and non-singular states.

\begin{acknowledgments}
The authors thank C.~E.~Zaspel, A.~Yu.~Galkin and A.~K.~Kolezhuk for fruitful discussions and help.  This work
was supported by INTAS, grant No.~97-31311 and partially by Volkswagen Stiftung, grant No.~I/75895.  One of us
(BI) thanks Montana State University for kind hospitality (NFS Grants No.~DMR-9974273 and No.~DMR-9972507).
\end{acknowledgments}

\end{document}